# Relativistic Solar Cells

*Paolo Umari,[a,b] Edoardo Mosconi,[c] Filippo De Angelis [c,*]*

[a] Dipartimento di Fisica e Astronomia, Università di Padova, via Marzolo 8, I-35131 Padova, Italy.

[b] CNR-IOM DEMOCRITOS, Theory@Elettra Group, c/o Sincrotrone Trieste, Area Science Park, Basovizza, I-34012 Trieste, Italy.

[c] Computational Laboratory for Hybrid/Organic Photovoltaics (CLHYO), CNR-ISTM, Via Elce di Sotto 8, I-06123, Perugia, Italy.

Hybrid **AMX$_3$ perovskites (A=Cs, CH$_3$NH$_3$; M=Sn, Pb; X=halide)** have revolutionized the scenario of emerging photovoltaic technologies.[1-7] Introduced in 2009 by Kojima *et al.*,[1] **a rapid evolution very recently led to 15% efficient solar cells.[2,5] CH$_3$NH$_3$PbI$_3$ has so far dominated the field, while the similar CH$_3$NH$_3$SnI$_3$ has not been explored for photovoltaic applications, despite the reduced band-gap.[8,9] Replacement of Pb by the more environment-friendly Sn would facilitate the large uptake of perovskite-based photovoltaics. Despite the extremely fast progress, the materials electronic properties which are key to the photovoltaic performance are relatively little understood. Here we develop an effective GW method incorporating spin-orbit coupling which allows us to accurately model the electronic, optical and transport properties of CH$_3$NH$_3$SnI$_3$ and CH$_3$NH$_3$PbI$_3$, opening the way to new materials design. The different CH$_3$NH$_3$SnI$_3$ and CH$_3$NH$_3$PbI$_3$ properties are discussed in light of their exploitation for solar cells, and found to be entirely due to relativistic effects.**

CH$_3$NH$_3$PbI$_3$, hereafter MAPbI$_3$ (MA=methylammonium), has been initially employed as sensitizing absorbing layer in conventional dye-sensitized solar cells (DSCs),[1] based on a mesoporous TiO$_2$ electron transporter, and later applied in solid-state DSCs based on a solid hole-transporter.[2] Lee *et al.*[3] have demonstrated that the chlorine-doped MAPb(I$_{1-x}$Cl$_x$)$_3$ perovskite can serve *both* as light-harvesting *and* electron conductor in meso-superstructured solar cells[3] employing an "inert" Al$_2$O$_3$ scaffold, reaching, together with a solid hole-transporter, a remarkable 12.3% conversion efficiency.[4] Liu *et al.* very recently reported a 15.4% efficient planar heterojunction solar cell, obtained by vapor deposition of MAPbI$_{3-x}$Cl$_x$ and by a solution-processed solid hole-transporter.[5] The high photovoltaic efficiency of these devices is mainly due to the optimal MAPbI$_3$/MAPbI$_{3-x}$Cl$_x$ perovskites band gap (~1.6 eV), which directly influences the solar cell photocurrent density (J$_{sc}$) and contributes to the open circuit voltage (V$_{oc}$) by setting the main solar cell energetics. A high carrier mobility within the perovskite ensures efficient collection of photo-generated charges.

Compared to MAPbI$_3$, the analogous MASnI$_3$ perovskite has been much less explored.[8,9] The two compounds show a similar tetragonal structure[9] (although in different temperature ranges) but different optical properties, with MASnI$_3$ (MAPbI$_3$) having an absorption onset at 1.2 (1.6) eV.[9,10] Experimental data also indicate that CsSnI$_3$ and MASnI$_3$ are excellent hole transporters,[6,9] while MAPb(I$_{1-x}$Cl$_x$)$_3$ and MAPbI$_3$ can sustain high rates of electron and hole transport, respectively.[3,7]

Understanding the origin of the different electronic properties of AMX$_3$ materials, with M=Sn and Pb, could represent a fundamental step towards the large-scale uptake of perovskites-based photovoltaics. In this context, a first principles computational approach capable of reliably calculating the materials band-gap and electronic/optical properties, thus trustfully allowing to design new materials and to interpret their properties, is fundamentally required. While standard Density Functional Theory (DFT) provides reliable structures and stabilities of perovskites,[11-13] it considerably underestimates the band-gap of these materials and in general of semiconductors. DFT with asymptotically correct functionals partly overcomes this shortcoming.[13] Many body perturbation theory, within the GW approach,[14,15] although more computationally demanding, constitutes a solid framework to improve upon DFT.[12,16] Contrary to expectations, DFT-calculated band-gaps of MAPbI$_3$ were in surprisingly good agreement, within ±0.1 eV, with experimental values.[17,18] For the supposedly similar ASnX$_3$ perovskites, DFT provided a ~1 eV band-gap underestimate.[8,11,12,19] Such an unbalanced description of Sn- and Pb-based materials hampers any predictive materials design/screening or comparative interpretation of their properties.

The large calculated band-gap difference between ASnX$_3$ and APbX$_3$ perovskites might be due to relativistic effects, particularly strong in Pb.[20,21] Relativistic effects are usually approximated by scalar relativistic (SR) and, to higher order, by spin-orbit coupling (SOC) contributions. A recent DFT investigation has confirmed a relevant SOC in MAPbX$_3$, leading to a strong, and opposite to the estimated GW correction, band-gap reduction.[22] This analysis poses the quest for a reliable and efficient theoretical framework for the simulation of ASnX$_3$ and APbX$_3$ perovskites and possibly of mixed Sn/Pb compounds. The method of choice is ideally a GW approach incorporating SOC.[23] A very effective GW implementation is also required, which was devised by some of us.[24] Here we develop a novel approach to introduce SOC effects into our efficient GW scheme. The resulting SOC-GW method is computationally affordable and it accurately reproduces the band-gap and electronic/optical properties of MASnI$_3$ and MAPbI$_3$.

Geometry optimization of the atomic positions (and cell parameters) of MAPbI$_3$ and MASnI$_3$ were performed by SOC-DFT (SR-DFT) in the tetragonal I4cm space group, with a unit cell made of four MAMI$_3$ units, containing 48 atoms and 200 electrons, see the MAPbI$_3$ structure in Figure 1. Using the experimental cell parameters, SR-DFT and SOC-DFT provide similar geometries, Supplementary Information. Cell relaxation leads to calculated bond lengths and lattice parameters in good agreement (within 1-2%) with experimental data, reproducing the long-short alternation of axial M-I bonds,[25] Figure 1. The expected shortening of M-I bonds upon Pb→Sn substitution is also nicely reproduced by our calculations, Supplementary Information.

The calculated band-gap values obtained at various levels of theory are graphically represented in Figure 2. All the investigated systems are characterized by a direct band-gap at the Γ point of the Brilluoin zone.[11,18] For MAPbI$_3$ and MASnI$_3$ the SR-DFT calculated band-gaps are 1.68 and 0.61 eV, to be compared to experimental values of 1.6 and 1.2 eV, respectively. Thus, while for MAPbI$_3$ the band-gap is reproduced by SR-DFT, for MASnI$_3$ a 0.6 eV band-gap underestimate is retrieved. Moving to SOC-DFT, the band-gap values are strongly underestimated, by as much as 1 eV, although a qualitatively correct band-gap variation is calculated, with a ~0.3 eV calculated difference against a ~0.4 eV experimental difference. The band-gap underestimate is in line with the expected behavior of DFT and with previous SOC-DFT results for MAPbI$_3$.[22] To correct the DFT-calculated band-gaps, we carried out SR- and SOC-GW calculations. SR-GW calculations for MAPbI$_3$ lead to a ~1 eV overestimate of the band-gap, while only ~0.3 eV band-gap overestimate is found for MASnI$_3$, again leading to an unbalanced description of the two systems, Figure 2. Rewardingly, SOC-GW delivers calculated band gaps (1.10 and 1.67 eV for MASnI$_3$ and MAPbI$_3$, respectively) in excellent agreement, within ±0.1 eV, with experimental values. Notice that ±0.1 eV is the inherent uncertainty of our calculations.

To make a direct connection between our calculations and solar cell operation, in Figure 2 we report the maximum J$_{sc}$ which can be extracted from a solar cell employing a material of varying band gap. The agreement between our SOC-GW calculated band-gaps and the experimental ones allows us to estimate the maximum J$_{sc}$ within ~10%. As an example, for MAPbI$_3$ we calculate a maximum J$_{sc}$ of ~25 mA/cm$^2$ against a ~28 mA/cm$^2$ value derived from the experimental band-gap. It is also worth noticing the potential of the MASnI$_3$ material to deliver extremely high J$_{sc}$ values due to its reduced band gap. This characteristic, along with its good transport properties, make this material highly promising to replace MAPbI$_3$, although some sensitivity of the material to the preparation conditions have been reported.[9]

Top J$_{sc}$ values measured for solar cells based on MAPbI$_3$ stand at ~21 mA/cm$^2$.[2,5] The reason for the non-optimal photocurrent generation can be traced back to the reduced light harvesting efficiency measured in the 600-800 nm range (~2.0–1.5 eV),[2] Figure 3. Based on our SOC-GW calculated electronic structure, we thus simulated the optical absorption spectrum of MAPbI$_3$, albeit neglecting electron-hole interactions, Supplementary Information. The employed procedure was shown to represent a reasonable approximation to the optical spectra of small band-gap semiconductors.[26] The results are reported in Figure 3, along with experimental data for MAPbI$_3$. The calculated data satisfactorily matches the experimental UV-vis spectrum: the band-gap absorption, the rise of the spectrum at higher energy and the feature at ~2.6 eV are nicely reproduced, despite the approximate spectral calculation. Compared to MAPbI$_3$, the absorption spectrum of MASnI$_3$ shows a red-shift (in line with the reduced band-gap) and increased intensity, Figure 3.

To provide a rationale for the observed band-gap and spectral variation, we investigated both structural and electronic factors. A SOC-GW calculation performed for MASnI$_3$ at the geometry and cell parameters of MAPbI$_3$ provided a band-gap of 1.48 eV, while rescaling the MAPbI$_3$ coordinates to the MASnI$_3$ cell parameters and substituting Pb by Sn, led to a SOC-DFT 0.17 eV band-gap increase compared to MASnI$_3$. This suggests that ~0.2 eV of the calculated band-gap difference (0.57 eV) are due to structural differences, such as the tilting of the MI$_6$ octahedra.

To investigate the electronic factors possibly responsible of the residual variations we analyze the GW-SOC Density of States (DOS) in Figure 4. A comparative picture of the electronic structure of the two systems can be gained by aligning the 2$p$ band of the CH$_3$NH$_3^+$ carbon atoms in the two materials, which appears as a narrow feature at ~8 eV below the valence band (VB) maximum in MASnI$_3$, Supplementary Information. This choice is justified by the fact that the organic molecules only weakly interact with the inorganic matrix by possible hydrogen bonding occurring through the ammonium groups. For the investigated systems the VB top is mainly composed by I $p$ orbitals, mixed in variable percentages with Pb or Sn $s$ orbitals, while the conduction band (CB) is mainly contributed by Pb or Sn $p$ orbitals, partly hybridized with I states.[17] The VB structure of the investigated systems is relatively similar, although MASnI$_3$ shows a widening and structuring of the VB compared to MAPbI$_3$ due to states found within ~1 eV below the VB maximum. The MAPbI$_3$ CB has a tail at lower energy compared to MASnI$_3$. Notably, in the absence of SOC the CB of MAPbI$_3$ has essentially the same structure as that of MASnI$_3$, Supplementary Information. The reduced spectral intensity calculated for MAPbI$_3$ appears thus to be due to the comparatively lower DOS close to the CB bottom, which is due to SOC. We can also compare the relative VB/CB position with available experimental data for MAPbI$_3$ and the analogous CsSnI$_3$ which indicate the VB and CB edges at 5.43-3.93 and 4.92-3.62 eV, respectively.[2, 6] Our calculations are in good agreement with experiments, and provide a ~0.6 eV VB energy downshift for MAPbI$_3$ compared to MASnI$_3$ along with a ~0.2 eV CB energy upshift.

The analysis of the aligned DOS allows us to understand the origin of the states responsible of the MASnI$_3$ reduced band gap, i.e. those states protruding out of the main VB peak, which are not found in MAPbI$_3$. These occupied states, of main I *p* character, have however a sizable Sn *s* contribution and are the result of the sizable anti-bonding interaction between Sn 5*s* and I 5*p* orbitals, Figure 4 and Supplementary Information. The corresponding Pb 6*s* orbitals are found at lower energy and have thus a lower tendency to mix with I 5*p* orbitals, thus the abrupt VB DOS rise found in MAPbI$_3$ compared to MASnI$_3$. Notice that the energetics of the 5*s*/6*s* shells in Sn/Pb are entirely due to relativistic effects, which substantially stabilize the Pb 6*s* shell leading to the so-called "inert 6*s*$^2$ lone pair".[20,21]

In line with the DOS changes, relativistic effects also deeply modify the band structure of the MAPbI$_3$ perovskite. Apart from the aforementioned band-gap change, introduction of SOC leads to an increased band dispersion along the investigated high symmetry directions of the Brilluoin zone. This leads to sizable differences for the calculated effective masses of electrons and holes, m$_e$ and m$_h$, respectively, reported in Table 2 and Supplementary Information. Considering average values, MASnI$_3$ is predicted to be a better hole transporter than MAPbI$_3$, while the two materials are predicted to show similar electron transport properties. These results, along with the analysis of the DOS width presented above, are in line with the experimental observations, whereby the CsSnI$_3$ perovskite was used as an efficient hole transporter in DSCs[6] and the MAPbI$_3$ and MAPb(I$_{1-x}$Cl$_x$)$_3$ compounds were found to efficiently transport both holes and electrons.[3,7] Our calculations also suggest MASnI$_3$ to be a potentially good electron transporter, in line with recent mobility results,[9] although to our knowledge this material has never been employed in solar cells. We can also compare the calculated reduced masses μ=m$_e$ • m$_h$/(m$_e$+m$_h$) with experimental data for MAPbI$_3$,[27] for which values of 0.09, 0.12 and 0.15 m$_0$ (m$_0$ is the electron mass) have been reported. Our minimum, average and maximum μ values calculated for MAPbI$_3$ are 0.08, 0.11 and 0.17 m$_0$, closely matching the experimental range of values.

In conclusion, we have devised a computationally efficient GW scheme incorporating SOC which has allowed us to unravel the electronic and optical properties of $MAPbI_3$ and $MASnI_3$ perovskites. The key to the different materials properties, thus to their photovoltaic performance, appears to be the different weight of relativistic effects in Sn- and Pb-based perovskites. This study provides the fundamental basis of understanding and the theoretical framework for the optimal exploitation of next generation "relativistic solar cells".

**AUTHORS CONTRIBUTION:**

PU conceived the work, implemented the SOC-GW code and performed the GW calculations. EM performed the DFT calculations and contributed to prepare the manuscript. FDA conceived the work and wrote the manuscript.

**ACKNOWLEDGMENT:**

The authors thank Dr. Annamaria Petrozza for providing us with the experimental absorption spectrum of $MAPbI_3$. We thank FP7-ENERGY-2010 Project ESCORT (contract No. 261023) and FP7-NMP-2009 Project SANS (contract No. 246124) for financial support.

**Method.**

We have extended the relativistic DFT scheme of Ref. [28], in which the spin-orbit coupling is included by 2-dimensional spinors and modeled by pseudopotentials, to our GW approach. Wavefunctions and charge densities are developed on a plane-waves basis sets. The two dimensional spinor exchange operator $\Sigma_x^{rel}$ is expressed as:

$$\Sigma_x^{rel}(r,\alpha;r',\alpha') = -e^2 \sum_{v=1,N_v^{rel}} \frac{\phi_v^{rel}(r,\alpha)\phi_v^{rel*}(r',\alpha')}{|r-r'|} \quad (1)$$

where the index $\alpha$ and $\alpha'$ run over the two spinor components of the $N_v^{rel}$ occupied relativistic KS states $\phi_v^{rel}$. For evaluating the self-energy $\Sigma_c^{rel}$ we have considered the suggestion of Ref. [23] of approximating the screened relativistic coulomb interaction $W_0^{c,rel}$ with that obtained from a scalar relativistic calculation $W_0^c$:

$$W_0^{c,rel}(r,\alpha;r',\alpha';\omega) \cong W_0^c(r,r';\omega)\delta_{\alpha,\alpha'} \quad (2)$$

For calculating the relativistic correlation part of the self-energy $\Sigma_c^{rel}$ we can calculate the DFT relativistic Green's function $G_0^{rel}$ considering explicitly only the lowest $N^{rel}$ relativistic states:

$$G_0^{rel}(r,\alpha;r',\alpha';\omega) \cong \sum_{i=1,N^{rel}} \frac{\phi_i^{rel}(r,\alpha)\phi_i^{rel*}(r',\alpha')}{\omega-\epsilon_i^{rel}} - \sum_{i=1,N^{rel}/2} \frac{\phi_i(r)\phi_i^*(r')}{\omega-\epsilon_i}\delta_{\alpha,\alpha'} + G_0(r,r';\omega)\delta_{\alpha,\alpha'}$$

$$(3)$$

where for simplicity in the scalar relativistic calculation we have considered doubly occupied states. In this way we still avoid sums over unoccupied KS states which would be particularly cumbersome when dealing with large model structures. The PBE exchange-correlation functional[29] and the Quantum Espresso program package[30] was used for all DFT calculations. Additional computational details for GW and DFT calculations are reported as Supplementary Information.

**Figures and Tables.**

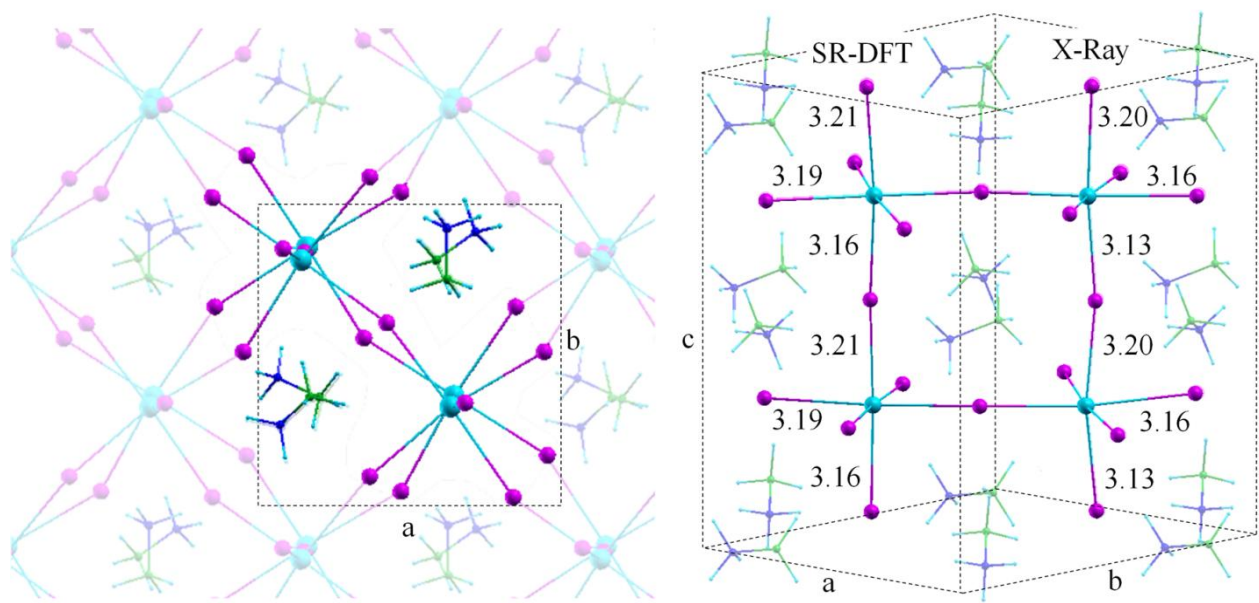

**Figure 1.** SR-DFT optimized structure of MAPbI$_3$ viewed from two different orientations. The unit cell is shown on the left. Calculated (average equatorial and axial values) and experimental[9] Pb-I distances (Å) are indicated on the right. Calculated (experimental [9,25]) cell parameters for MAPbI$_3$: a=8.78 (8.85-8.86) Å; c= 12.70 (12.64-12.66) Å. Calculated (experimental [8, 9]) cell parameters for MASnI$_3$: a=8.71 (8.76-8.73) Å; c= 12.46 (12.43-12.50) Å. Pb=light blue; I=purple; N=blue; C= green.

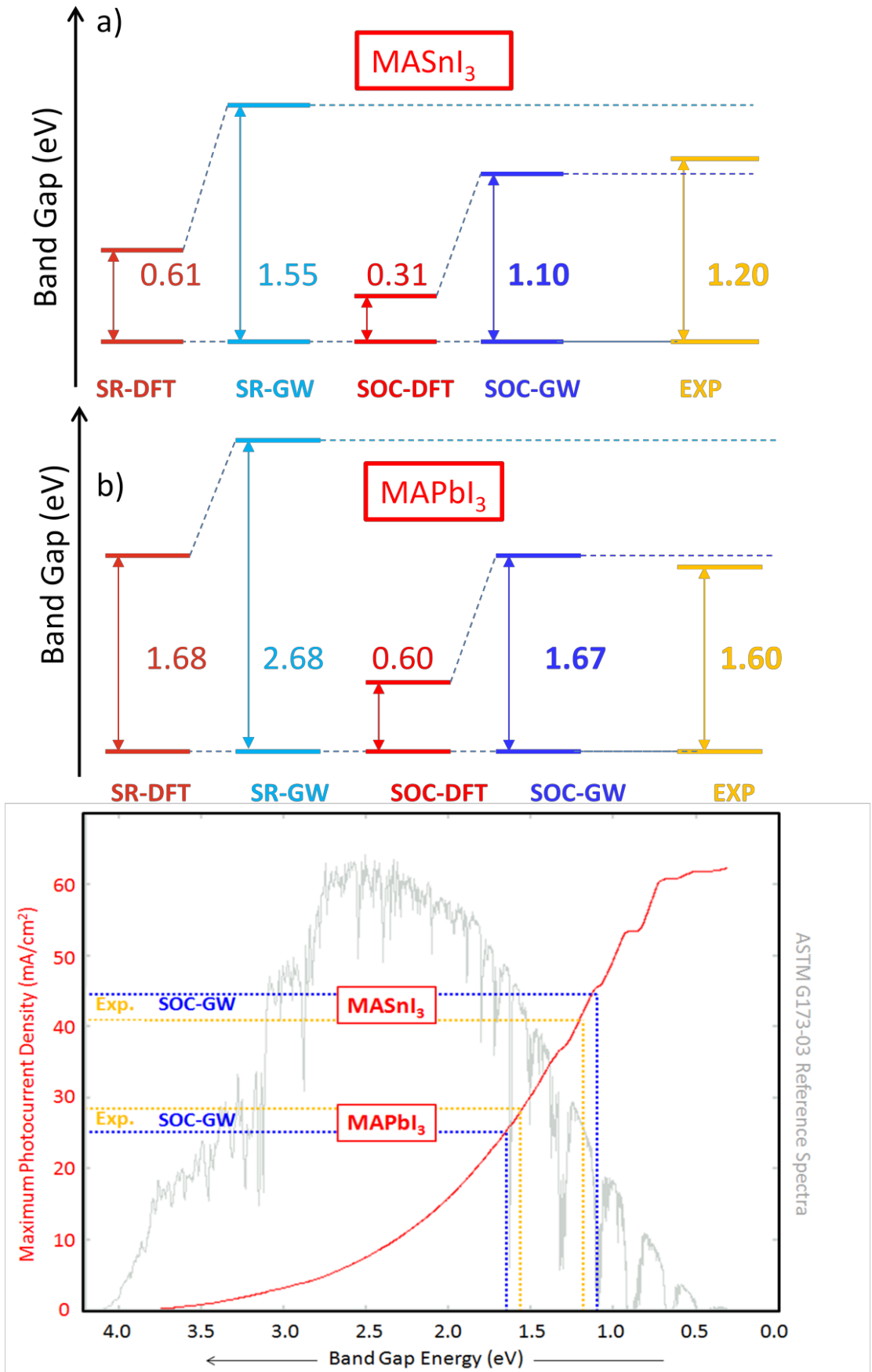

**Figure 2.** Calculated band-gaps at various levels of theory for MASnI$_3$ (a) and MAPbI$_3$ (b) perovskites. (c) Maximum short-circuit photocurrent density which can be extracted from a solar cell employing MASnI$_3$ and MAPbI$_3$, as obtained by integration of ASTM G173-03 reference spectrum with the assumption of 100% IPCE above the band-gap. Experimental data from Ref. [1,27] and [9].

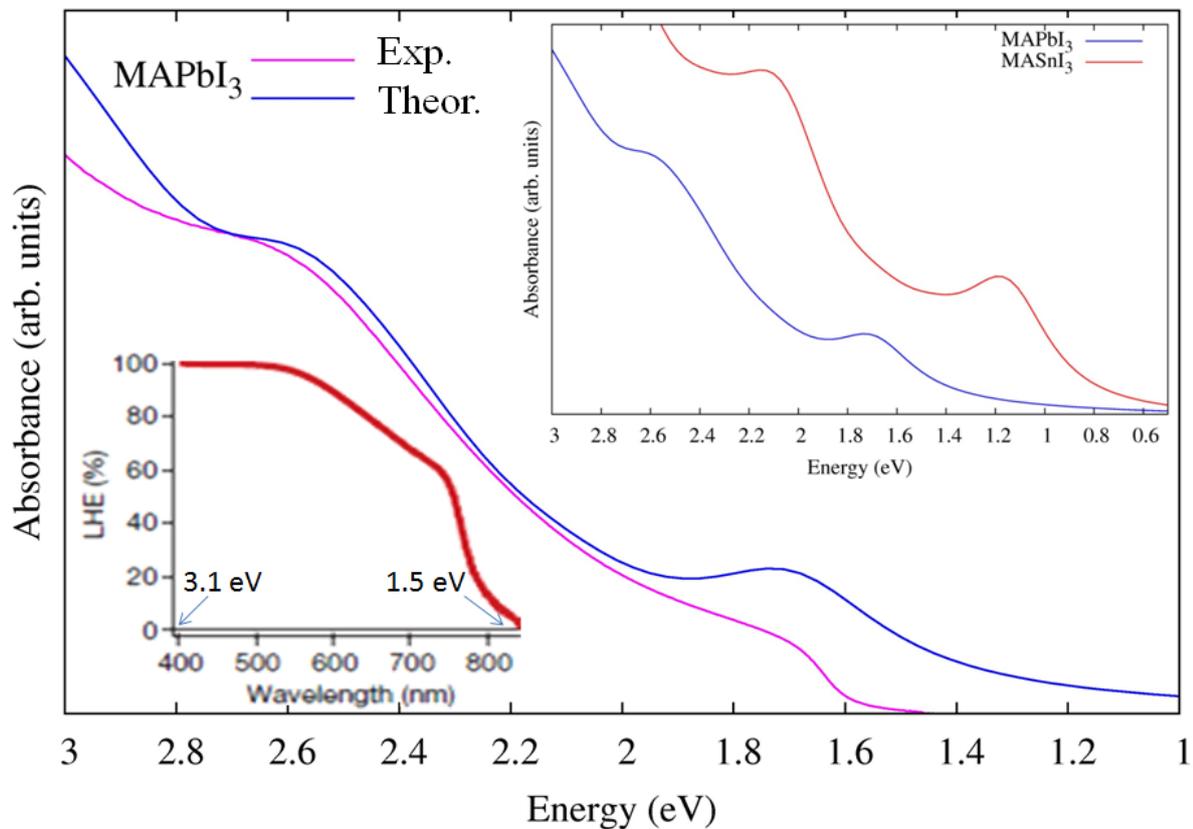

**Figure 3**. Comparison between the experimental UV-vis spectrum of MAPbI$_3$ (red line) and the SOC-GW calculated one (blue line). Notice that the experimental spectrum has been scaled to match the intensity of the calculated one in correspondence of the high energy feature. Top right inset: Comparison between the SOC-GW calculated spectra of MASnI$_3$ (red line) and MAPbI$_3$ (blue line). Bottom left inset: LHE for the 15% MAPbI$_3$-based solar cell of Ref. [2]. The experimental spectrum was recorded for at room temperature for MAPbI$_3$ casted on a mesoporous TiO$_2$ film.

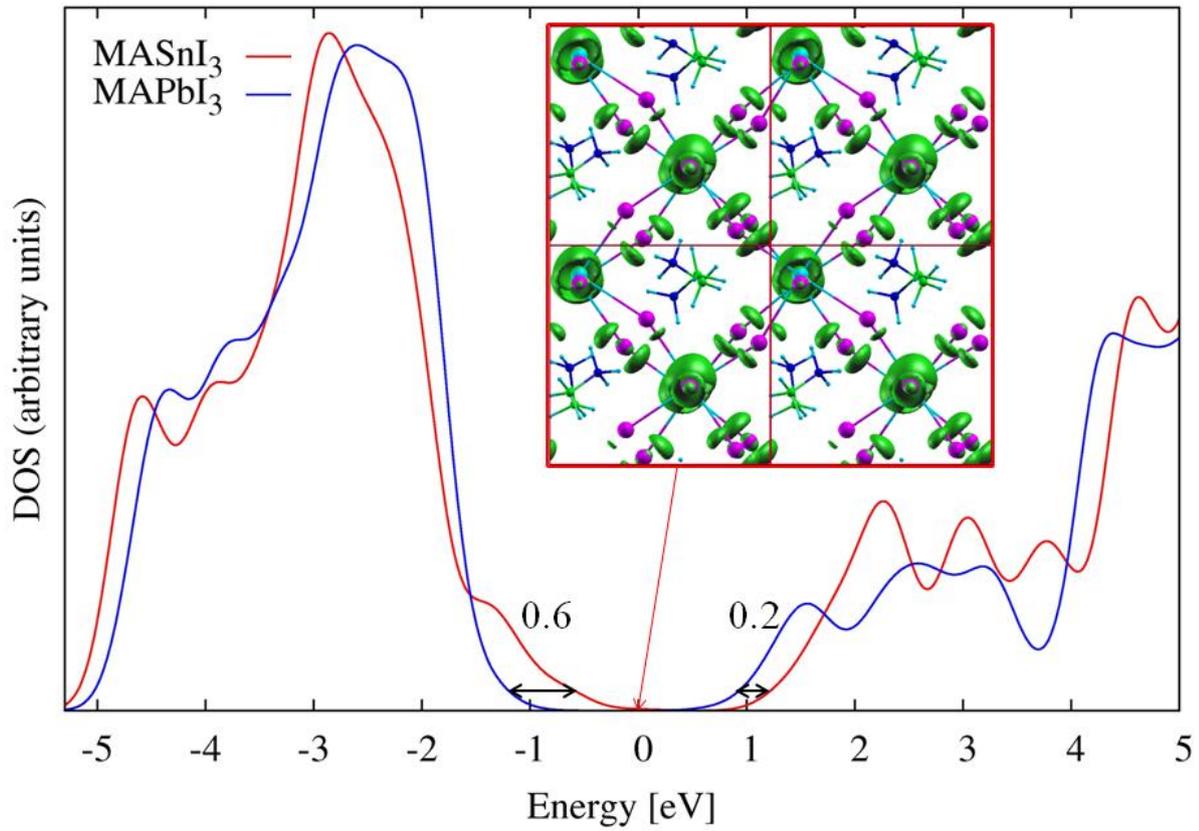

**Figure 4**. Electronic DOS for MASnI$_3$ (red line) and MAPbI$_3$ (blue line) calculated by SOC-GW. The DOS have been aligned at the carbon 2$p$ peak. Inset: SOC-DFT isodensity plot of the highest occupied state of MASnI$_3$ calculated at the Γ point.

**Table 1**. SOC-GW effective masses for holes ($m_h$) and electrons ($m_e$) calculated by parabolic fitting of the VB and CB (alpha manifold) along the directions Γ (0,0,0) →M (0.5, 0.5,0); Γ→Z (0,0,0.5); Γ→X (0,0.5,0); Γ→A (0.5,0.5,0.5); Γ→R (0,0.5,0.5) and the corresponding reduced masses (μ) for MAPbI$_3$ and MASnI$_3$.

|  | MAPbI$_3$ | | | MASnI$_3$ | | |
|---|---|---|---|---|---|---|
|  | $m_h$ | $m_e$ | μ | $m_h$ | $m_e$ | μ |
| Γ→M | 0.34 | 0.15 | 0.10 | 0.18 | 0.21 | 0.10 |
| Γ→Z | 0.51 | 0.25 | 0.17 | 0.29 | 0.31 | 0.15 |
| Γ→X | 0.26 | 0.14 | 0.09 | 0.15 | 0.18 | 0.08 |
| Γ→A | 0.28 | 0.13 | 0.09 | 0.16 | 0.14 | 0.07 |
| Γ→R | 0.22 | 0.13 | 0.08 | 0.14 | 0.12 | 0.06 |
| AVG | 0.32 | 0.16 | 0.11 | 0.18 | 0.19 | 0.09 |